\documentclass[12pt]{article}
\usepackage{amsmath}
\usepackage{amssymb}

\newcommand{\veff}{V_{\rm{eff}}}
\newcommand{\psia}{\psi^{(\alpha)}}
\newcommand{\ea}{E^{(\alpha)}}
\newcommand{\na}{{\cal N}^{(\alpha)}}
\newcommand{\lambdaa}{\frac{\lambda}{\alpha}}
\newcommand{\kta}{\tilde{K}^{(\alpha)}}
\newcommand{\ka}{K^{(\alpha)}}
\newcommand{\N}{\mathbb{N}}
\newcommand{\Aa}{A^{(\alpha)}}
\newcommand{\ca}{C^{(\alpha)}}

\title{
Quadratic algebras and position-dependent mass Schr\"odinger equations}
\author{C Quesne\thanks{Electronic mail: cquesne@ulb.ac.be}\\ 
{\small\sl Physique Nucl\'eaire Th\'eorique et Physique Math\'ematique,  Universit\'e Libre de Bruxelles,} \\ 
{\small\sl Campus de la Plaine CP229, Boulevard~du Triomphe, B-1050 Brussels, Belgium}}
\date{ }
\begin{document}
\baselineskip=22pt plus 1pt minus 1pt
\maketitle
\begin{abstract}
During recent years, exact solutions of position-dependent mass Schr\"odinger equations have inspired intense research activities, based on the use of point canonical transformations, Lie algebraic methods or supersymmetric quantum mechanical techniques. Here we highlight the interest of another approach to such problems, relying on quadratic algebras. We illustrate this point by constructing spectrum generating algebras for a class of $d$-dimensional radial harmonic oscillators with $d\ge2$ (including the one-dimensional oscillator on the line via some minor changes) and a specific mass choice. This provides us with a counterpart of the well-known su(1,1) Lie algebraic approach to the constant-mass oscillators.
\end{abstract}

\section{Introduction}
During recent years, quantum mechanical systems with a position-dependent (effective) mass (PDM) have attracted a lot of attention and inspired intense research activites (see \cite{cq07a} for a list of references). They are indeed very useful in the study of many physical systems, such as electronic properties of semiconductors and quantum dots, nuclei, quantum liquids, $^3$He clusters, metal clusters, etc. Furthermore, the PDM presence in quantum mechanical problems may reflect some other unconventional effects, such as a deformation of the canonical commutation relations or a curvature of the underlying space. Hermitian PDM Hamiltonians may also be equivalent to some $\cal PT$-symmetric systems with constant mass at lowest order of perturbation theory.\par
%
%
To find exact solutions for some PDM Schr\"odinger equations, use has been made of point canonical transformations, Lie algebras or supersymmetric quantum mechanical techniques by extending methods known for constant mass. Very recently, nonlinear algebras, especially quadratic ones, have started to be employed. The presence of one of them has been signalled in a one-dimensional PDM problem \cite{roy}. Then the quadratic algebra generated by the integrals of motion of a two-dimensional superintegrable PDM system has been studied \cite{cq07b}. Finally, a quadratic algebra approach has allowed the construction of spectrum generating algebras for a class of $d$-dimensional radial harmonic oscillators with specific PDM. In the present contribution, we briefly summarize the results of the last work (for more details, see \cite{cq07a}).\par
%
%
\section{\boldmath Constant-mass $d$-dimensional radial harmonic oscillator}
In units wherein $\hbar = 1$ and the mass $m_0 = 1/2$, the radial Schr\"odinger equation for the $d$-dimensional harmonic oscillator ($d \ge 2$) can be written as
\begin{equation}
  \left(- \frac{d^2}{dr^2} + \frac{L(L+1)}{r^2} + \frac{1}{4} \omega^2 r^2\right) \psi(r) = E 
  \psi(r),  \label{eq:HO}
\end{equation}
where $r$ runs on the half-line $0 < r < \infty$, $L$ is defined by $L = l + (d-3)/2$ in terms of the angular momentum quantum number $l$ and the radial wavefunction is actually $r^{-(d-1)/2} \psi(r)$.
\par
%
%
Equation (\ref{eq:HO}) has an infinite number of bound-state solutions
\begin{equation}
  \psi_{n,L}(r) = {\cal N}_{n,L} r^{L+1} L_n^{(L+\frac{1}{2})}(\tfrac{1}{2} \omega r^2) 
  e^{-\frac{1}{4} \omega r^2}, \qquad n = 0, 1, 2, \ldots,  \label{eq:HO-psi}
\end{equation}
corresponding to the energy eigenvalues
\begin{equation}
  E_{n,L} = \omega (2n + L + \tfrac{3}{2}).  \label{eq:HO-E}
\end{equation}
In (\ref{eq:HO-psi}), $L_n^{(\alpha)}(y)$ denotes a Laguerre polynomial and ${\cal N}_{n,L}$ is a normalization coefficient.\par
%
%
All the wavefunctions (\ref{eq:HO-psi}), corresponding to a given value of $L$ and $n=0$, 1, 2,~\ldots, belong to a single positive-discrete series unitary irreducible representation $D^+_k$ of an su(1,1) Lie algebra. The latter is generated by the operators
\begin{equation}
\begin{split}
  K_0 &= \frac{1}{2\omega} \left(- \frac{d^2}{dr^2} + \frac{L(L+1)}{r^2} + \frac{1}{4} \omega^2 r^2\right)
    , \\
  K_{\pm} &= \frac{1}{2\omega} \left[\frac{d^2}{dr^2} - \frac{L(L+1)}{r^2} + \frac{1}{4} \omega^2 r^2
    \mp \omega \left(r \frac{d}{dr} + \frac{1}{2}\right)\right],  
\end{split} \label{eq:HO-gen}
\end{equation}
satisfying the commutation relations $[K_0, K_{\pm}] = \pm K_{\pm}$, $[K_+, K_-] = - 2K_0$, and the Hermiticity properties $K_0^{\dagger} = K_0$, $K_{\pm}^{\dagger} = K_{\mp}$, while its Casimir operator reads $C = - K_+ K_- + K_0 (K_0 - 1)$. The lowest weight characterizing the irreducible representation is here
$k = \frac{1}{2} \left(L + \frac{3}{2}\right)$.\par
%
%
The wavefunctions $\psi_{n,L}(r)$, $n=0$, 1, 2,~\ldots, are simultaneous eigenfunctions of $C$ and $K_0$,
\begin{equation}
\begin{split}
  C \psi_{n,L}(r) &= k(k-1) \psi_{n,L}(r) = \tfrac{1}{4} \left(L + \tfrac{3}{2}\right)\left(L - \tfrac{1}{2}\right)
    \psi_{n,L}(r), \\
  K_0 \psi_{n,L}(r) &= \mu \psi_{n,L}(r) = (k+n) \psi_{n,L}(r) = \frac{1}{2\omega} E_{n,L} \psi_{n,L}(r).
\end{split}  \label{eq:HO-action}
\end{equation}
Furthermore, $K_+$ and $K_-$ act on them as
\begin{equation}
\begin{split}
  K_+ \psi_{n,L}(r) &= [(\mu-k+1)(\mu+k)]^{1/2} \psi_{n+1,L}(r) = \left[(n+1) \left(n+L+\tfrac{3}{2}\right)
    \right]^{1/2} \psi_{n+1,L}(r), \\
  K_- \psi_{n,L}(r) &= [(\mu-k)(\mu+k-1)]^{1/2} \psi_{n-1,L}(r) = \left[n \left(n+L+\tfrac{1}{2}\right)
    \right]^{1/2} \psi_{n-1,L}(r).
\end{split}  \label{eq:HO-action-bis}
\end{equation}
This shows that su(1,1) is a spectrum generating algebra for the $d$-dimensional radial harmonic oscillator.\par
%
%
\section{\boldmath PDM $d$-dimensional radial harmonic oscillator}
Let us now consider a PDM $d$-dimensional harmonic oscillator, whose radial Schr\"odinger equation is obtained by replacing in (\ref{eq:HO}) the radial momentum $p_r = - {\rm i} d/dr$ by some deformed one, $\pi_r = \sqrt{f(\alpha; r)}\, p_r \sqrt{f(\alpha; r)}$, where $f(\alpha; r) = 1 + \alpha r^2$ and $\alpha$ is a positive real constant. The result of this substitution reads
\begin{equation}
  \left(\pi_r^2 + \frac{L(L+1)}{r^2} + \frac{1}{4} \omega^2 r^2\right) \psia(r) = \ea \psia(r),  
  \label{eq:PDM-HO}
\end{equation}
which is equivalent to 
\begin{equation}
  \left(- \frac{d}{dr} \frac{1}{M(\alpha; r)} \frac{d}{dr} + \veff(\alpha;r)\right) \psia(r) =\ea \psia(r),  
  \label{eq:radial-PDM} 
\end{equation}
where $M(\alpha; r)$ and $\veff(\alpha;r)$ denote the PDM  
\begin{equation}
  M(\alpha; r) = \frac{1}{f^2(\alpha; r)} = \frac{1}{(1 + \alpha r^2)^2}
\end{equation}
and the effective potential
\begin{equation}
  \veff(\alpha;r) = \frac{L(L+1)}{r^2} + \frac{1}{4} (\omega^2 - 8 \alpha^2) r^2 - \alpha,
\end{equation}
respectively.\par
%
%
It is known \cite{bagchi} that equation (\ref{eq:PDM-HO}) has an infinite number of bound states giving rise to a quadratic energy spectum
\begin{equation}
 \ea_{n,L} = \alpha \left(4n^2 + 4n(L+1) + L + 1 + (4n + 2L + 3) \frac{\lambda}{\alpha}\right), \qquad n=0, 
 1, 2, \ldots,  \label{eq:PDM-E}
\end{equation}
where $\lambda = \frac{1}{2}(\alpha + \Delta)$ and $\Delta = \sqrt{\omega^2 + \alpha^2}$. Furthermore, the corresponding eigenfunctions can be written as 
\begin{equation}
  \psia_{n,L}(r) = \na_{n,L} r^{L+1} P_n^{\left(\lambdaa - \frac{1}{2}, L + \frac{1}{2}\right)}(t)
  f^{-[\lambda + (L+2) \alpha]/(2\alpha)},  \label{eq:PDM-psi-bis}
\end{equation}
where $P_n^{\left(\lambdaa - \frac{1}{2}, L + \frac{1}{2}\right)}(t)$ is a Jacobi polynomial in the variable $t = 1 - 2/f = (-1 + \alpha r^2)/(1 + \alpha r^2)$ and $\na_{n,L}$ is some normalization coefficient.\par
%
%
The Hamiltonian in the PDM Schr\"odinger equation (\ref{eq:PDM-HO}), the variable $t$ appearing in the wavefunctions (\ref{eq:PDM-psi-bis}) and their commutator can be taken as the generators
\begin{equation}
  \kta_1 = \pi_r^2 + \frac{L(L+1)}{r^2} + \frac{1}{4} \omega^2 r^2, \qquad \kta_2 = t, \qquad \kta_3 = 
  - 4{\rm i}{\alpha} \left(2 \frac{r}{f} \pi_r + {\rm i}t\right)  \label{eq:PDM-gen}
\end{equation}
of a quadratic algebra, which can be identified with a quantum Jacobi algebra QJ(3) \cite{granovskii}. Their commutation relations are indeed given by
\begin{equation}
\begin{split}
  \bigl[\kta_1, \kta_2\bigr] &= \kta_3, \\
  \bigl[\kta_2, \kta_3\bigr] &= 8 \alpha \bigl(1 - \tilde{K}^{(\alpha)2}_2\bigr), \\
  \bigl[\kta_3, \kta_1\bigr] &= - 8 \alpha \bigl\{\kta_1, \kta_2\bigr\} - 16 \alpha^2 \left[\lambdaa 
    \left(\lambdaa - 1\right) + L(L+1) - 1\right] \kta_2 \\
  & \quad - 16 \alpha^2 \left[\lambdaa \left(\lambdaa - 1\right) - L(L+1)\right].
\end{split}  \label{eq:PDM-com}
\end{equation}
It can be shown that this algebra has a single positive-discrete series unitary irreducible representation $D^+_{p_0}$ with $p_0 = \frac{1}{2} \left(\lambdaa + L\right)$ related to the lowest energy eigenvalue $\ea_{0,L}$.\par
%
%
It is possible to construct another basis $\bigl(\ka_0, \ka_+, \ka_-\bigr)$ of this quadratic algebra, satisfying the following three properties:
\begin{itemize}
\item[(i)] $\ka_0$ is proportional to the Hamiltonian  of the problem, while $\ka_+$ (resp.\ $\ka_-$) is a raising (resp.\ lowering) ladder operator, which means that, up to some multiplicative factor, it transforms $\psia_{n,L}$ into $\psia_{n+1,L}$ (resp.\ $\psia_{n-1,L}$) for any $n \in \N$ (resp.\ $n \in \N^+$) with the additional condition that $\ka_-$ annihilates $\psia_{0,L}$.
\item[(ii)] The operators $\ka_0$, $\ka_+$, $\ka_-$ satisfy the same Hermiticity properties as $K_0$, $K_+$, $K_-$, i.e., $K^{(\alpha)\dagger}_0 = \ka_0$ and $K^{(\alpha)\dagger}_{\pm} = \ka_{\mp}$.
\item[(iii)] In the $\alpha \to 0$ limit, they go over to the su(1,1) generators $K_0$, $K_+$, $K_-$.
\end{itemize}
Such properties explicitly show (i) that the PDM $d$-dimensional radial harmonic oscillator Schr\"odinger equation (\ref{eq:PDM-HO}) admits a spectrum generating algebra, and (ii) that the latter is a deformed su(1,1) algebra.
\par
%
%
The operators $\ka_0$, $\ka_+$, $\ka_-$ can indeed be expressed in terms of $\kta_1$, $\kta_2$, $\kta_3$ as
\begin{equation}
\begin{split}
  \ka_0 &= \frac{1}{4\lambda} \kta_1,\\
  \ka_{\pm} &= \pm \frac{1}{16\lambda} \Aa_{\pm} (\delta \pm 1) \sqrt{\frac{\delta \pm 2}{\delta}} = 
      \pm \frac{1}{16\lambda} (\delta \mp 1) \sqrt{\frac{\delta}{\delta \mp 2}} \Aa_{\pm},
\end{split}
\end{equation}
where
\begin{equation}
\begin{split}
  \Aa_{\pm} &= \kta_3 - 4\alpha \kta_2 (1 \mp \delta) + 4\alpha \frac{\left(\lambdaa-L-1\right) 
       \left( \lambdaa+L\right)}{1 \pm \delta}, \\
  \delta &= \sqrt{\frac{\kta_1}{\alpha} + \lambdaa \left(\lambdaa-1\right) + L(L+1)}, 
\end{split}
\end{equation}
and their action on $\psia_{n,L}(r)$ is given by
\begin{equation}
\begin{split}
  \ka_0 \psia_{n,L} &= \frac{1}{4\lambda} \ea_{n,L} \psia_{n,L}, \\
  \ka_+ \psia_{n,L} &= \frac{\alpha}{\lambda} \left[(n+1) \left(n+L+\frac{3}{2}\right) \left(n+\lambdaa+L+1  
    \right) \left(n+\lambdaa+\frac{1}{2}\right)\right]^{1/2} \psia_{n+1,L}, \\
  \ka_- \psia_{n,L} &= \frac{\alpha}{\lambda} \left[n \left(n+L+\frac{1}{2}\right) \left(n+\lambdaa+L\right) 
    \left(n+\lambdaa-\frac{1}{2}\right)\right]^{1/2} \psia_{n-1,L}.
\end{split}  
\end{equation}
\par
%
%
The three deformed su(1,1) generators $\ka_0$, $\ka_+$ and $\ka_-$ satisfy the commutation relations
\begin{equation}
\begin{split}
  \bigl[\ka_0, \ka_{\pm}\bigr] &= \pm \frac{\alpha}{\lambda} \ka_{\pm} (\delta \pm 1) = \pm 
  \frac{\alpha}{\lambda} (\delta \mp 1) \ka_{\pm}, \\
  \bigl[\ka_+, \ka_-\bigr] &= - \frac{\alpha\delta}{\lambda} \left(2 \ka_0 + \frac{\alpha}{4\lambda}\right),   
\end{split}
\end{equation}
and the algebra admits a Casimir operator
\begin{equation}
  \ca = - \ka_+ \ka_- + K^{(\alpha)2}_0 - \frac{\alpha}{\lambda} \left(\delta - \frac{5}{4}\right) \ka_0 -
  \frac{\alpha^2}{8\lambda^2} \delta, 
\end{equation}
with eigenvalues given in
\begin{equation}
  \ca \psia_{n,L} = \left[\frac{1}{4} \left(1 - \frac{\alpha}{\lambda}\right) \left(L + \frac{3}{2}\right)
  \left(L - \frac{1}{2}\right) - \frac{3\alpha^2}{16\lambda^2} L(L+1)\right] \psia_{n,L}.  
\end{equation}
\par
%
%
\section{Final remarks}
We have shown that a quadratic algebra approach provides us with a useful tool for constructing spectrum generating algebras for a class of $d$-dimensional radial harmonic oscillators with specific PDM depending on some parameter $\alpha$. The well-known su(1,1) algebra corresponding to the constant-mass limit is retrieved for $\alpha \to 0$. The case of the one-dimensional harmonic oscillator on the full line (either with a constant mass or with a similar PDM) can be derived from that of the radial oscillator by formally replacing $L$ by $-1$ or 0. The two resulting unitary irreducible representations of the spectrum generating algebra are spanned by even- and odd-parity wavefunctions, respectively.\par
%
%

\end{document}